\RequirePackage{silence}
\documentclass[fleqn,usenatbib,useAMS]{mnras} 
\usepackage{graphicx}
\let\normalbigoplus \bigoplus
\usepackage{amsmath}
\let\bigoplus \normalbigoplus

\usepackage{amsfonts}
\usepackage{float}
\usepackage{bm}
\usepackage[perpage,symbol*]{footmisc}
\usepackage{ae,aecompl}
\usepackage{array}
\usepackage{soul}
\usepackage{mathtools}
\usepackage{multirow}

\title{Scale-invariant dynamics in the Solar System} 

\author[Indranil Banik \& Pavel Kroupa]{Indranil Banik$^{1}$\thanks{Email:
\href{mailto:ibanik@astro.uni-bonn.de}{ibanik@astro.uni-bonn.de} (Indranil Banik)\newline $~~~~~~~~~~~~~~~~~~~~$ \href{mailto:pavel@astro.uni-bonn.de}{pkroupa@uni-bonn.de} (Pavel Kroupa)} and Pavel Kroupa$^{1,2}$\\
$^{1}$Helmholtz-Institut f\"ur Strahlen und Kernphysik (HISKP), University of Bonn, Nussallee 14$-$16, D-53115 Bonn, Germany \\
$^{2}$Charles University, Astronomical Institute, Faculty of Mathematics and Physics, V Hole\v{s}ovi\v{c}k\'ach 2, CZ-180 00 Praha 8, Czech Republic}

\pubyear{2020}
\begin{document}
\label{firstpage}
\pagerange{\pageref{firstpage}--\pageref{lastpage}}

\maketitle

\begin{abstract}

The covariant scale-invariant dynamics (SID) theory has recently been proposed as a possible explanation for the observed dynamical discrepancies in galaxies (Maeder \& Gueorguiev 2020). SID implies that these discrepancies $-$ commonly attributed to dark matter $-$ arise instead from a non-standard velocity-dependent force that causes two-body near-Keplerian orbits to expand. We show that the predicted expansion of the Earth-Moon orbit is incompatible with lunar laser ranging data at $>200\sigma$. Moreover, SID predicts that the gravitating mass of any object was much smaller in the past. If true, a low-mass red giant star must be significantly older than in standard theory. This would make it much older than the conventional age of the Universe, which however is expected to be similarly old in SID. Moreover, it is not completely clear whether SID truly contains new physics beyond General Relativity, with several previous works arguing that the extra degree of freedom is purely mathematical. We conclude that the SID model is falsified at high significance by observations across a range of scales, even if it is theoretically well formulated.

\end{abstract}


\section{Introduction}
\label{Introduction}

The Large Synoptic Survey Telescope \citep{Ivezic_2019} has recently been renamed the Vera C. Rubin Observatory in honour of her observational work establishing what is still one of the great mysteries of astronomy \citep{Rubin_Ford_1970}. Their work revealed very large dynamical discrepancies between the actual rotation curves of galaxies and the predictions of Newtonian gravity applied to their luminous matter distributions, as also shown by several other authors \citep[e.g.][]{Babcock_1939, Rogstad_1972}. Such acceleration discrepancies are usually attributed to dark matter haloes surrounding each galaxy \citep{Ostriker_Peebles_1973}. The dark matter would have to consist of hypothetical particles not in the well-tested standard model of particle physics. This interpretation is challenged by continued null detection of any dark matter particles in sensitive searches, for instance in 11 years of Fermi data on dwarf spheroidal satellites of the Milky Way \citep{Hoof_2020} and in sensitive terrestrial experiments \citep{Liu_2017}.

Regardless of the hypothetical particle physics that might give rise to dark matter, assuming it holds galaxies together leads to inconsistencies on a variety of scales ranging from hundreds to millions of parsecs \citep{Kroupa_2012, Kroupa_2015}. The longest studied problem is that the galactic acceleration discrepancies follow some remarkable regularities \citep{Famaey_McGaugh_2012} that can be summarized as a unique relation between the acceleration $g$ inferred from the rotation curve and the Newtonian gravitational field $g_{_N}$ generated by the baryonic distribution \citep{McGaugh_Lelli_2016}. Such a radial acceleration relation (RAR) was predicted several decades earlier using Modified Newtonian Dynamics \citep[MOND,][]{Milgrom_1983}. In this model, the dynamical effects usually attributed to dark matter are instead provided by an acceleration dependence of the gravity law $-$ the gravitational field strength $g$ at distance $r$ from an isolated point mass $M$ transitions from the Newtonian $GM/r^2$ law to
\begin{eqnarray}
	g ~=~ \frac{\sqrt{GMa_{_0}}}{r} ~~~\text{for } ~ g, g_{_N} \ll a_{_0} \, .
	\label{Deep_MOND_limit}
\end{eqnarray}
MOND (or Milgromian dynamics) introduces $a_{_0}$ as a fundamental acceleration scale of nature below which the deviation from Newtonian dynamics becomes significant. Empirically, $a_{_0} = 1.2 \times {10}^{-10}$ m/s$^2$ to match galaxy rotation curves \citep{Begeman_1991, McGaugh_2011}.

Remarkably, this is approximately where the classical energy density of a gravitational field \citep[equation 9 of][]{Peters_1981} becomes comparable to the dark energy density $u_{_\Lambda} \equiv \rho_{_\Lambda} c^2$ that conventionally explains the accelerating expansion of the Universe \citep{Efstathiou_1990, Ostriker_Steinhardt_1995, Riess_1998, Perlmutter_1999}.
\begin{eqnarray}
	\frac{g^2}{8\mathrm{\pi}G} ~<~ u_{_\Lambda} ~~\Leftrightarrow~~ g ~\la~ 2\mathrm{\pi}a_{_0} \, .
	\label{MOND_quantum_link}
\end{eqnarray}
In regions of space with such a weak gravitational field, the dominant contribution to the energy density is $u_{_\Lambda}$, which is often identified with the quantum-mechanical zero point energy density of the vacuum. If this is correct, poorly understood quantum gravity effects could well be rather important to such regions $-$ but these are totally neglected in General Relativity. MOND could be an empirical way to include such effects, as suggested by the coincidence of scales in Equation \ref{MOND_quantum_link} \citep[e.g.][]{Milgrom_1999, Pazy_2013, Verlinde_2016, Smolin_2017}.

Regardless of its underlying microphysical explanation, MOND can accurately match the rotation curves of a wide variety of spiral galaxies across a vast range in mass, surface brightness, and gas fraction using only the distribution of luminous matter \citep[figure 5 of][]{Lelli_2017}. Fits to individual rotation curves show that intrinsic scatter about its predictions must be ${<13\%}$ and is consistent with 0 \citep{Li_2018}. A few discrepant galaxies were claimed by \citet{Rodrigues_2018}, but it was later shown that the actual discrepancies are either very mild or arise when the distance is particularly uncertain and could plausibly be outside the range they allow \citep{Kroupa_2018}. MOND can also explain the rotation curves of elliptical galaxies, at least when these can be measured in a sub-dominant rotating disc of neutral hydrogen \citep[figure 8 of][]{Lelli_2017}.

The successes of MOND extend beyond near-circular orbital motion in the non-relativistic regime. A relativistic version of MOND has recently been developed in which gravitational waves propagate at the speed of light \citep{Skordis_2019}, consistent with the near-simultaneous detection of the gravitational wave event GW170817 and its electromagnetic counterpart \citep{LIGO_Virgo_2017}.  Galactic globular clusters can be used to test MOND in dispersion-dominated systems \citep{Baumgardt_2005, Haghi_2011}. In this regard, NGC 2419 was claimed to be problematic for MOND \citep{Ibata_2011A, Ibata_2011B}. It was later shown that the disagreement could be caused by plausible systematic uncertainties like a mild departure from spherical symmetry, rotation within the sky plane, or a radially-dependent polytropic index \citep{Sanders_2012A, Sanders_2012B}. In MOND, we also expect NGC 2419 to be affected by the `external field effect' (EFE) whereby the internal gravity binding a system is weakened when it is embedded in an external gravitational field, a consequence of the non-linear MOND equations \citep{Milgrom_1986}. However, including the Galactic external field did not much improve the agreement with observations \citep{Derakshani_2014}. MOND can also explain the velocity dispersion of the ultra-diffuse galaxy Dragonfly 2 \citep[DF2,][]{Kroupa_2018_Nature}, DF4 \citep{Haghi_2019}, and DF44 \citep{Bilek_2019}, with the latter galaxy having a measured dispersion profile that is consistent with MOND at ${2.40 \sigma}$ \citep{Haghi_2019_DF44}. The calculations for DF2 and DF4 are significantly affected by the EFE. DF44 is more isolated than the almost gas-free DF2 \citep{Chowdhury_2019, Sardone_2019} and indeed has a higher internal velocity dispersion despite a similar baryonic distribution. In a MOND context, the EFE was recently confirmed at high significance based on the relative velocities of wide binary stars in the Solar neighbourhood $-$ these are inconsistent with MOND without the Galactic EFE \citep{Pittordis_2019}. The more relevant case with the EFE is quite similar to the Newtonian case, though a more careful analysis could distinguish them in the near future \citep{Banik_2018_Centauri, Banik_2019}. Recent progress with hydrodynamic MOND simulations indicates that it can naturally form exponential disc galaxies out of a collapsing gas cloud \citep{Wittenburg_2020} and produce realistic morphologies for the interacting Antennae galaxies \citep{Renaud_2016}.

Most of these observations only became apparent long after the MOND field equation was first published \citep{Bekenstein_Milgrom_1984}, making these achievements successful a priori predictions. It is difficult to explain the success of these predictions in a conventional gravity context, even with the observational facts in hand \citep{Desmond_2016, Desmond_2017, Ghari_2019}. In particular, section 4.2 of the latter work showed that it is still rather difficult to get a tight RAR despite a diversity of rotation curve shapes at fixed peak velocity \citep{Oman_2015}.

Despite its successes, the theoretical underpinnings of MOND remain unclear. An important clue may be that in the deep-MOND limit, the dynamics of gravitational systems become scale invariant $-$ scaling the spacetime co-ordinates by some factor $\lambda$ yields another valid solution to the equations of motion \citep{Milgrom_2009_DML}. In principle, $\lambda$ could be time-dependent. This is at the basis of the fully relativistic scale invariant dynamics theory \citep[SID,][]{Maeder_1978, Maeder_1979}. SID applies the Weyl Integrable Geometry \citep{Canuto_1977} and reproduces some of the MOND phenomenology, in particular a tight RAR \citep{Maeder_2020}. There are some differences in its predicted form when $g_{_N} \la 0.01 a_{_0}-$ while MOND predicts that $g \to \sqrt{g_{_N} a_{_0}}$, SID predicts that $g$ flattens out at $\approx 0.1 a_{_0}$ (see their figure 1). This leads to potentially testable consequences in low surface brightness dwarf galaxies.

In this contribution, we avoid discussing possible theoretical issues with SID, which may in fact be identical to classical General Relativity as the extra degree of freedom may be a mathematical artefact \citep{Tsamis_1986, Jackiw_2015}. We assume that there is genuinely new physics in SID and focus on its high-acceleration limit, where SID has some unusual consequences.

After introducing the SID model and its context (Section \ref{Introduction}), we discuss how it can be constrained by Solar System ephemerides (Section \ref{Solar_System_ephemerides}). We also consider what SID implies for stellar evolution (Section \ref{Stellar_evolution}). Our conclusions are given in Section \ref{Conclusions}.


\section{Solar System ephemerides}
\label{Solar_System_ephemerides}

SID has some success reproducing the RAR \citep[section 5 of][]{Maeder_2020}. This comes about because of a non-standard velocity-dependent force which causes a slow outwards expansion of a two-body near-Keplerian orbit (see their equation 26). This has always been an important part of the SID theory \citep{Maeder_1978}. As shown in equation 68 of \citet{Maeder_1979}, the extra term causes a near-circular orbit of radius $r$ to expand as
\begin{eqnarray}
	\dot{r}_{_{SID}} ~=~ \frac{r}{t} \, ,
	\label{r_dot_SID}
\end{eqnarray}
where $t$ is the time since the Big Bang and $\dot{q} \equiv \frac{dq}{dt}$ for any quantity $q$. The $_{SID}$ subscript indicates a theoretical expectation. In this contribution, it will be important to measure time from the Big Bang since the dynamical equations of SID are not invariant under a translation of the time variable.

Since the expansion rate of the SID universe is similar to the Lambda-Cold Dark Matter ($\Lambda$CDM) standard cosmological paradigm \citep[section 4.1 of][]{Maeder_2020}, we assume that currently ${t = 13.8}$~Gyr \citep{Planck_2015}. Thus, the 9.58 AU orbit of Saturn should expand at $\dot{r}_{_{SID}} = 104$~m/yr. Over a decade, the expansion should easily be detectable in Cassini radio tracking data given that its accuracy is $\approx 32$~m \citep{Viswanathan_2017}. However, they found no significant deviation of Saturn from its conventionally calculated trajectory. It is difficult to see how a constant outwards expansion could be masked by e.g. changing the masses of other planets.

\subsection{Lunar laser ranging}
\label{LLR}

Applying Equation \ref{r_dot_SID} to the Earth-Moon distance of ${r = 3.84 \times 10^8}$~m, we get that $\dot{r}_{_{SID}} = 28$~mm/yr. \citet{Maeder_1979} predicted that this effect should be observable in ``sufficiently accurate observations with laser reflectors'' (see their section 7). Nowadays, the Earth-Moon distance is constrained to mm accuracy by lunar laser ranging \citep[LLR,][]{Adelberger_2017}. The Moon is indeed receding from the Earth, but at a rate of $\dot{r}_{obs} = {38.05 \pm 0.04}$~mm/yr \citep[section 3c of][]{Folkner_2014}. This is caused by the tides it raises in Earth's oceans. These tidal bulges are carried ahead of the sub-lunar point because Earth rotates much faster than the Moon orbits it. \citet{Folkner_2014} estimated that the effect of these tides is known to an accuracy of $\approx 0.5\%$ or 0.2 mm/yr, making this the limit to any unconventional contributions to $\dot{r}$.

Nonetheless, we consider the possibility that this calculation is seriously in error such that the tidal contribution to $\dot{r}_{obs}$ is only $\dot{r}_{tide} = 10$~mm/yr, with SID contributing the remaining 28 mm/yr in line with Equation \ref{r_dot_SID}. The total $\dot{r}$ would then be consistent with the $\dot{r}_{obs}$ measured by LLR.

An important aspect of the SID contribution is that orbital velocities $v$ do not change even though the radius $r \propto t$ (Equation \ref{r_dot_SID}). Since $v^2 = GM/r$ for a circular orbit around any object with mass $M$, its gravitational parameter $GM \propto t$ in this model.\footnote{This partially cancels the orbit expansion due to the non-standard velocity-dependent force, which would otherwise be twice as much.} Thus, in the absence of tides, SID predicts that the Moon's orbital angular frequency $\Omega \propto t^{-1}$ \citep[equation 71 of][]{Maeder_1979}. However, if the lunar orbit expands at the same rate due to tides in a conventional context, Kepler's Third Law implies that $\Omega \propto t^{-3/2}$. Combining the tidal and SID contributions, we get that
\begin{eqnarray}
    \frac{\dot{\Omega}}{\Omega} ~=~ -\frac{\dot{r}_{_{SID}} + \frac{3}{2}\dot{r}_{tide}}{r} \, .
    \label{Omega_dot}
\end{eqnarray}

\begin{table}
	\centering
	\begin{tabular}{cccc}
		\hline
		\raisebox{-0.002\textheight}{Model} & \raisebox{-0.002\textheight}{$\dot{r}_{tide}$, mm/yr} & \raisebox{-0.002\textheight}{$\dot{r}_{_{SID}}$, mm/yr} & \raisebox{-0.002\textheight}{$\dot{\Omega}$, $\arcsec$/century\textsuperscript{2}} \\
		\hline
		Standard & 38 & 0 & $-25.74$ \\
		SID & 28 & 10 & $-19.42$ \\
		Observed & \multicolumn{2}{c}{Total $= 38.05 \pm 0.04$} & $-25.80 \pm 0.03$ \\
		\hline
	\end{tabular}
	\caption{The observed lunar recession rate is partly caused by tides raised on the Earth. It might contain an additional term predicted by SID, but if so, the effect of tides must be reduced since the total is tightly constrained by lunar laser ranging \citep{Folkner_2014}. Therefore, the scenarios predict different angular deceleration rates for the lunar orbit (Equation \ref{Omega_dot}). The observed rate comes from their section 3c.}
	\label{Omega_dot_table}
\end{table}

The conventional expectation is that $\dot{r}_{_{SID}} = 0$ and $\dot{r}_{tide} = 38$~mm/yr, implying that $\dot{\Omega} = -25.74 \arcsec$/century\textsuperscript{2}. If instead the precisely observed $\dot{r}_{obs}$ consists of $\dot{r}_{tide} = 10$~mm/yr and $\dot{r}_{_{SID}} = 28$~mm/yr, we should observe that $\dot{\Omega} = -19.42 \arcsec$/century\textsuperscript{2}. These possibilities are summarized in Table \ref{Omega_dot_table}.

Space age observations of the Moon tell us that $\dot{\Omega} = -25.80 \pm 0.03 \arcsec$/century\textsuperscript{2} \citep[section 3c of][]{Folkner_2014}. This could be affected by oscillatory perturbations from other planets, but fortunately a much longer baseline is available if we also consider pre-telescopic records of Solar and lunar eclipses over the past 2700 years \citep{Stephenson_1995}. This yields an estimated $\dot{\Omega}$ of ${-26 \pm 1}\arcsec$/century\textsuperscript{2}, with the error derived from their estimated timing uncertainty of 0.9 s/century\textsuperscript{2} (their section 5a) and the fact that changing $\dot{\Omega}$ by $0.5 \arcsec$/century\textsuperscript{2} affects the eclipse timings by 0.46 s/century\textsuperscript{2} (their section 2b). Impressively, this section of their work found that $\dot{\Omega} = -26.0 \arcsec$/century\textsuperscript{2} fits the ancient records much better than $\dot{\Omega} = -26.2 \arcsec$/century\textsuperscript{2}, correctly anticipating subsequent refinements to the modern value of $\dot{\Omega}$.

Reconciling this estimate with SID requires changing $\dot{\Omega}$ by $\Delta \dot{\Omega} = 6 \arcsec$/century\textsuperscript{2}. The SID-predicted angular deceleration of the Earth is only $0.94 \arcsec$/century\textsuperscript{2}, so SID-induced changes in its orbital period have little effect on this discrepancy. Over ${T = 2000}$ years, the discrepancy amounts to a shift in the angular position of the Moon by $T^2 \Delta \dot{\Omega}/2 = 1200 \arcsec$. Since the Moon takes 27.3 days to orbit the Earth, it would take 36 minutes to rotate through ${1200 \arcsec}$. Ancient astronomers were well capable of noticing such a large time difference, especially since some eclipses occurred close to sunrise or sunset. Although there is some degeneracy between $\dot{\Omega}$ and changes in the Earth's rotation rate, the time of day provides a tight constraint on the latter. The very fact that an eclipse occurred tightly constrains the lunar orbit while the positions of background stars constrain Earth's orbit around the Sun. In this way, careful observations and record-keeping can break the various degeneracies involved. Therefore, the SID theory is falsified at extremely high significance by both modern and ancient observations.

\section{Time-varying $G$ and stellar evolution}
\label{Stellar_evolution}

We have seen that SID predicts the gravitational parameter $GM \propto t$. If this is caused by changes in $M$, then the masses of fundamental particles would need to grow. In particular, changing the electron mass $\propto t$ would cause a similar change in the energy of hydrogen spectral lines, i.e. the Rydberg constant would change. For much heavier elements, the expected shift is not exactly the same due to special relativistic corrections that become more important for atoms orbiting a more charged nucleus. This allowed \citet{Karshenboim_2008} to place stringent constraints on possible time evolution of the Rydberg constant, completely ruling out the possibility that it grows $\propto t$.

Thus, the SID model requires $G \propto t$, implying a present value of $\dot{G}/G = 7.2 \times 10^{-11}$/yr. This very slow change is in significant tension with the precision pulsar timing constraint from PSR J0437-4715 \citep[$\dot{G}/G = \left( -0.5 \pm 1.8 \right) \times 10^{-11}$/yr,][]{Verbeist_2008}.

Over a Hubble time, the much larger expected change in $G$ would significantly alter stellar evolution since the core pressure and temperature depend sensitively on $G$. As a result, its expected variation is in strong tension with astroseismic observations of the ancient star KIC 7970740 \citep[$\dot{G}/G = \left( 1.2 \pm 2.6 \right) \times 10^{-12}$/yr,][]{Bellinger_2019}.

More generally, since SID does not much change the age of the universe, it would require significant revision to our understanding of main sequence turnoffs in the oldest globular clusters \citep{Vandenberg_2013, Correnti_2018}. We consider this unlikely because the nuclear processes in stars are rather well understood and the turnoff stars have a mass of ${\approx 0.8 M_\odot}$, making them only slightly less massive than the Sun.

Any time variation of $G$ would also change the Chandrasekhar mass $\propto G^{-3/2}$ \citep{Gaztanaga_2002}. This would affect the luminosities of Type Ia supernovae, allowing observations of them to set limits on $\dot{G}$. Based on this idea, the analysis of \citet{Mould_2013} constrained $\dot{G}/G$ to the range $\left(-3, 7.3 \right) \times 10^{-11}$/yr, marginally consistent with the SID expectation of $7.2 \times 10^{-11}$/yr. Future improvements to this technique could yield more stringent constraints.

\section{Conclusions}
\label{Conclusions}

The SID cosmological model claims to successfully reproduce the acceleration-dependent pattern of dynamical discrepancies in galaxies \citep[the RAR,][]{Maeder_2020}. This is due to an extra term in the equations, which in the Solar System implies that the Moon's orbit around the Earth should expand at a rate of $\dot{r}_{_{SID}} = 28$~mm/yr (Section \ref{LLR}). Such an expansion of two-body orbits has always been part of SID \citep{Maeder_1978, Maeder_1979}, with the latter work explicitly suggesting that this prediction should be tested with laser ranging. The observed lunar recession rate of $\dot{r}_{obs} = {38.05 \pm 0.04}$~mm/yr \citep{Folkner_2014} can be accounted for by tidal dissipation in Earth's oceans. For the SID theory to be correct, there must be an extremely large error in our understanding of how this process works, even though the uncertainty should be $\la 0.2$~mm/yr. Supposing nonetheless that our understanding of terrestrial tides is significantly in error, $\dot{r}_{obs}$ would have to consist of 28 mm/yr from SID and 10 mm/yr from tides. This would cause the angular frequency of the lunar orbit to decrease by $19.42 \arcsec$/century\textsuperscript{2}, contradicting the observed rate of $25.80 \pm 0.03 \arcsec$/century\textsuperscript{2}. This rate has remained stable for several thousand years and is well explained using standard mechanics (Table \ref{Omega_dot_table}).

Looking beyond the Earth-Moon system, SID predicts that the orbit of Saturn expands at 104 m/yr (Section \ref{Solar_System_ephemerides}). This probably violates constraints from the Cassini mission since the ranging accuracy is ${\approx 32}$ m and the orbiter functioned for over a decade \citep[table 11 of][]{Viswanathan_2017}. Unlike oscillatory perturbations from other planets, SID predicts a constant outwards expansion that is difficult to mask by e.g. changing the mass of Jupiter.

In addition to predicting that orbits expand, SID also implies significant time variation of any object's gravitational parameter $GM$ (Section \ref{Stellar_evolution}). Since changing the masses of fundamental particles would violate precise laboratory constraints, the most plausible interpretation is that the change occurs solely in the gravitational constant $G$. This would substantially affect stellar evolution because any ancient star would have fused hydrogen at a much lower rate in the early universe, implying low-mass red giants must be significantly older than in standard theory. However, the SID model also implies the age of the universe is similar to that in $\Lambda$CDM \citep[section 4.1 of][]{Maeder_2020}.

In addition to these observational difficulties, it is not completely clear that SID is in fact distinct from classical General Relativity $-$ the extra degree of freedom may be a mathematical artefact without any physical effects \citep{Tsamis_1986, Jackiw_2015}. In any case, SID is unable to satisfy Solar System constraints and contradicts stellar astrophysics to a substantial extent. Therefore, some other way should be found to explain the impressive successes of Milgromian dynamics in a covariant framework. The recently proposed relativistic MOND theory of \citet{Skordis_2019} may be a significant step in this direction.

\section*{Acknowledgements}

IB is supported by an Alexander von Humboldt postdoctoral research fellowship. The authors are grateful to Andre Maeder for carefully presenting his work at the Bonn Gravity2019 conference. They also thank Richard Woodard and Moti Milgrom for helpful discussions, and the referee for comments which helped to clarify this letter.

\section*{Data availability}

The data underlying this article are available in the article.

\bibliographystyle{mnras}
\bibliography{SID_bbl}
\bsp
\label{lastpage}
\end{document}